\documentstyle[11pt,epsfig]{article}

\begin{document}

\begin{titlepage}
\begin{center}
\large STATE RESEARCH CENTER OF RUSSIA\\
\Large INSTITUTE FOR HIGH ENERGY PHYSICS\\
\end{center}

\bigskip

\begin{flushright}
{ IHEP 98-48\\}
{physics/9809037\\}
\end{flushright}     

\bigskip

\begin{center}{\Large ON OBSERVABILITY OF SIGNAL OVER BACKGROUND}
\end{center}

\bigskip

  \begin{center}
{\large
    S.I.~Bityukov$^1$ (IHEP, Protvino RU-142284, Russia),\\
    N.V.~Krasnikov$^2$ (INR, Moscow 117312, Russia)
}
\end {center}

\vspace{2cm}
  
\begin{abstract}
Several statistics used by physicists 
to declare the signal observability over the background 
are compared.  
It is shown that the frequentist method of testing a precise hypothesis 
allows one to estimate the power value of criteria with specified
level of significance for the considered statistics by  
Monte Carlo calculations. The application of this approach for
the analysis of discovery potential of experiments is discussed. 
\end{abstract}

\vspace{2cm}

\small
\noindent
\rule{3cm}{0.5pt}\\
$^1$E-mails: bityukov@mx.ihep.su,~~Serguei.Bitioukov$@$cern.ch\\
$^2$E-mails: krasniko@ms2.inr.ac.ru,~~Nikolai.Krasnikov$@$cern.ch\\

\vspace{1cm}

\begin{center}
Protvino 1998
\end{center}

\end{titlepage}

\newpage

\begin{center}
{\large \bf Introduction}
\end{center}

One of the common tasks for searching experiments is the detection of
a predicted new Phenomenon. As a rule the estimations of an expected mean $N_s$ 
for the signal events of new Phenomenon  and  
$N_b$ for the background events are known. Then we want to know is the given 
experiment able to detect new Phenomenon or not. 
To check the statement about the observation of Phenomenon 
a researcher uses some function of the observed number of events 
-- a statistic. The value of this statistic for detected $x$ events allows one 
to find the degree of confidence of the discovery statement. After 
having drawn a conclusion on the
observation of Phenomenon, two possibilities for mistake are available:
to state that Phenomenon is absent but in real life it exists
(Type I error), or to state that Phenomenon exists but it is absent
(Type II error).

In this paper we compare the ``signal significances'' used 
by the researchers for the hypothesis testing about 
the observation of Phenomenon:

\begin{itemize}
\item[(a)]
``significance'' $S_1 = \displaystyle \frac{N_s}{\sqrt{N_b}}$~\cite{1},
\item[(b)]
``significance'' $S_2 = \displaystyle \frac{N_s}{\sqrt{N_s + N_b}}$~\cite{2,3},
\item[(c)]
``significance'' 
$S_{12} = \displaystyle \sqrt{N_s + N_b} - \sqrt{N_b}$~\cite{4},

\item[(d)]
likelihood ratio as is defined in references~\cite{5,6}.
\end{itemize}

For this purpose we formulate the null and alternative hypotheses, construct 
the statistical test, determine the rejection region by Monte Carlo
calculations, make the decision and find the power of test for 
the criteria with a specified level of significance. 
We also use an equal-tailed test
to study the behaviour of Type I and Type II errors versus
$N_s$ and $N_b$ for specified values of $S_1$ and $S_2$.
The hypotheses testing results obtained by Monte Carlo calculations 
are compared with result obtained by the direct calculations of
probability density functions.  

\section{Notations}

Let us study a physical process during a fixed time.
The estimations of the average number of signal events which indicate
new Phenomenon ($N_s$) and of the average number of background events ($N_b$) 
in the experiment are given. We suppose that the events have the
Poisson distributions with the parameters $N_s$ and $N_b$, 
i.e. the random variable $\xi \sim Pois(N_s)$
describes the signal events and the random variable $\eta \sim Pois(N_b)$ 
describes the background events. 
Say we observed $x$ events -- the realization  
of the studying process $X = \xi + \eta$
($x$ is the sum of signal and background events in the experiment). 
Here $N_s$, $N_b$ are non-negative real numbers and $x$ is an integer.
The classical frequentist methods of testing a precise hypothesis allow one 
to construct a rejection region and determine associated error probabilities
for the following ``simple'' hypotheses:

$H_0:~X \sim Pois(N_s + N_b)$ versus $H_1:~X \sim Pois(N_b)$,
where $Pois(N_s + N_b)$ and $Pois(N_b)$ have the  probability
density functions (p.d.f.'s) \\
$f_0(x) = \displaystyle \frac{(N_s+N_b)^x}{x!} e^{-(N_s+N_b)}$
for the case of presence and   
$f_1(x) = \displaystyle \frac{(N_b)^x}{x!} e^{-(N_b)}$ \\
for the case of absence of signal events in the universe population.

In Fig.1 the p.d.f.'s $f_0(x)$~(a) and $f_1(x)$~(b) for the case 
$N_s+N_b=104$ and $N_b=53$ (\cite{3}, Table.13, cut 6) are shown. 
As is seen the intersection of these p.d.f.'s takes place. 
Let us denote the threshold (critical value) that divides 
the abscissa in Fig.1 into the rejection region and the area of accepted
hypothesis $H_0$ via $N_{ev}$. The incorrect rejection of the null hypothesis $H_0$,
the Type I error (the statement that Phenomenon is absent, but it is present), 
has the probability $\displaystyle \alpha = \sum_{x=0}^{N_{ev}}{f_0(x)}$, 
and the incorrect acceptance 
of $H_0$, the Type II error (the statement that Phenomenon exists, 
but it is absent), has the probability
$\displaystyle \beta = \sum_{x=N_{ev}+1}^{\infty}{f_1(x)}$. 
The dependence of $\alpha$ and $\beta$ on the value of $N_{ev}$ 
for above example is presented in Fig.2. 

\begin{figure}[ht]
\epsfig{file=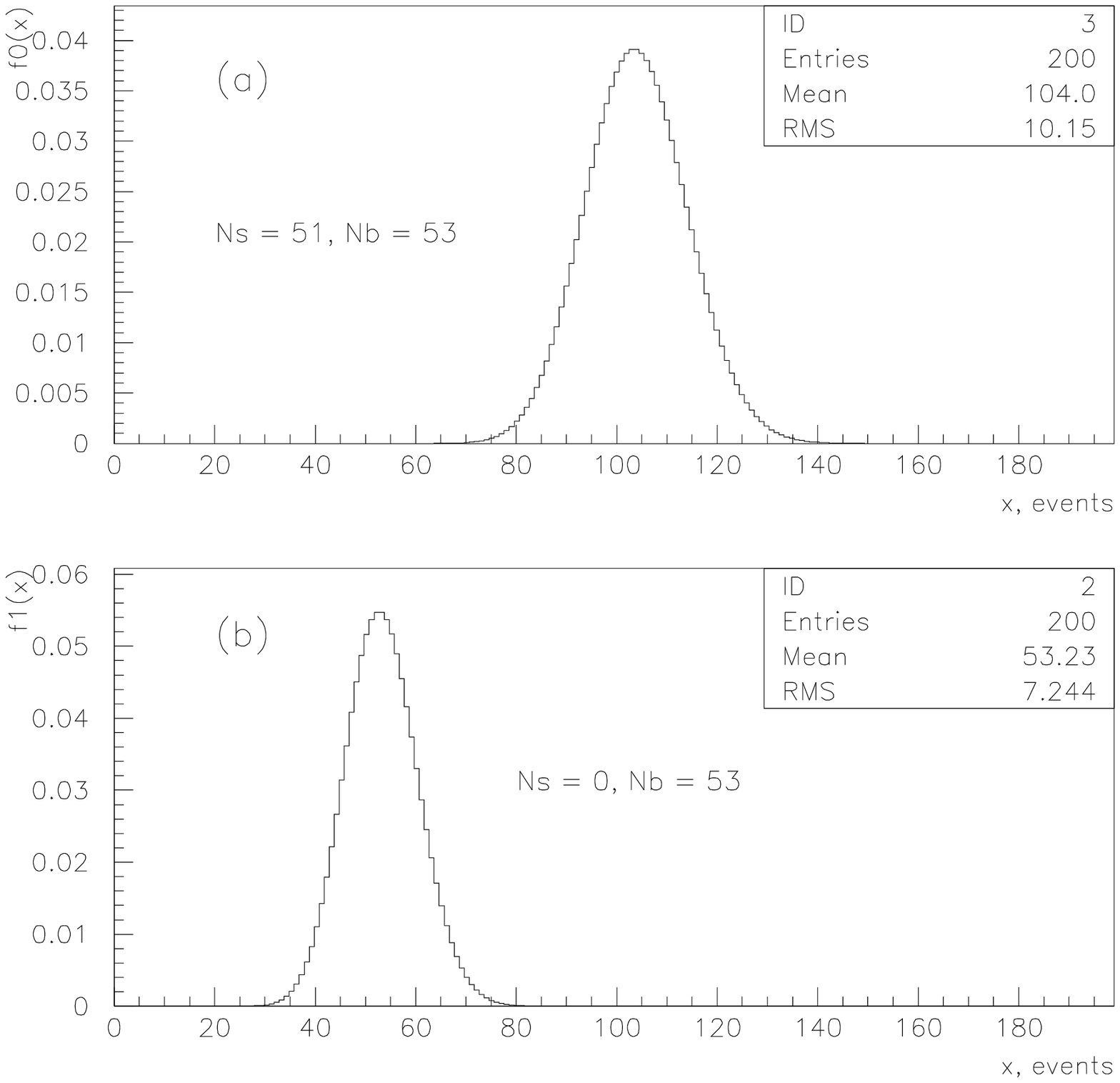,width=15cm}
\caption{The probability density functions
$f_0(x)$~(a) and $f_1(x)$~(b) for the case of 51 signal events 
and 53 background events obtained by direct calculations
of the probabilities.}
\label{fig.1}
\end{figure}
        
\begin{figure}[ht]
\epsfig{file=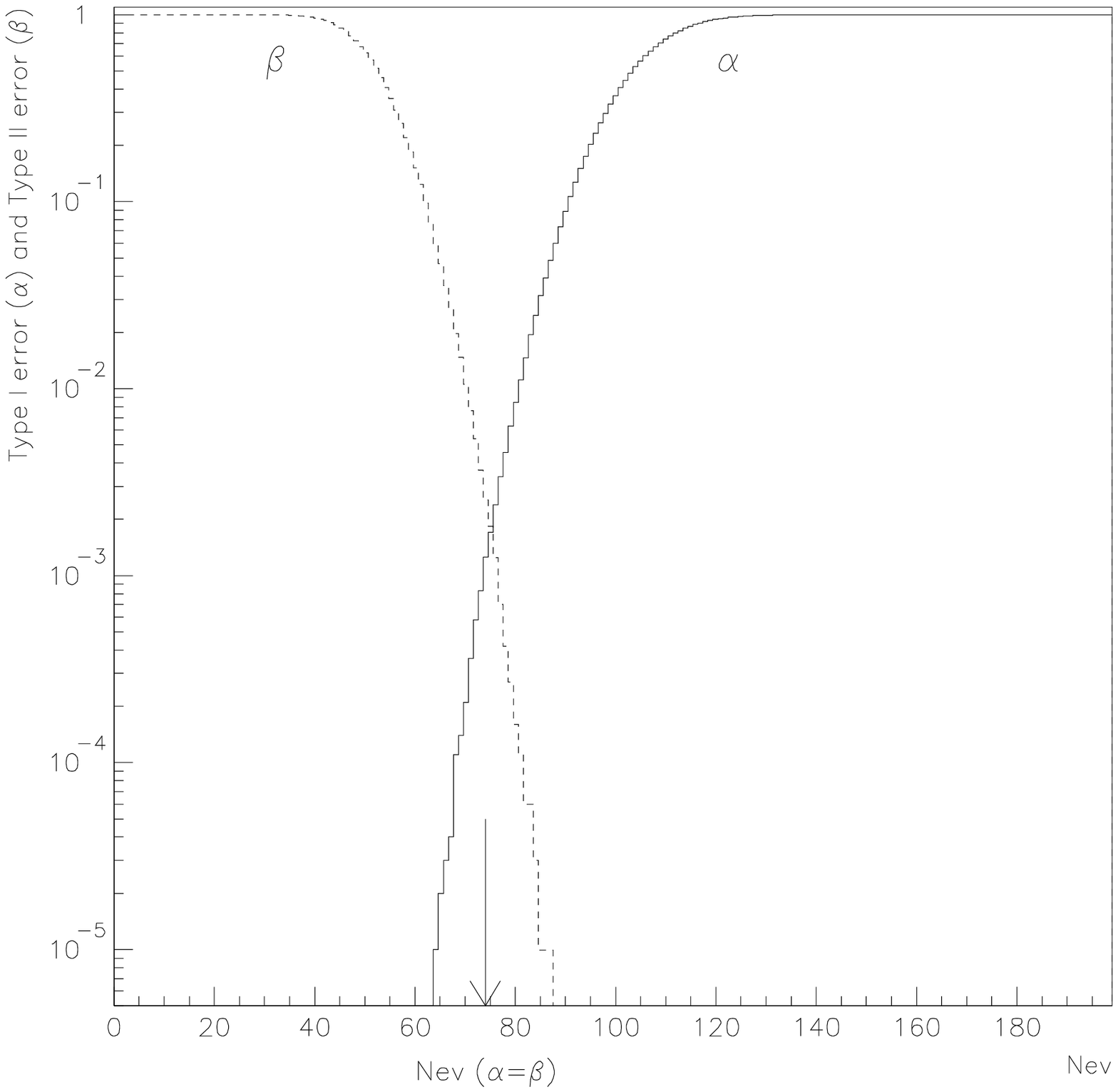,width=15cm}
\caption{The dependence of Type I $\alpha$ and Type II $\beta$ 
errors on $N_{ev}$ for the case of 51 signal events 
and 53 background events.}
\label{fig.2}
\end{figure}

\section{Hypothesis testing}

In this Section we show the procedure of the rejection
region construction  for the likelihood ratio~\cite{5}. 

We denote by $B(x) = \displaystyle \frac{f_0(x)}{f_1(x)}$ the likelihood ratio
of $H_0$ to $H_1$ in the area of existing $B(X)$. 
The decision to either reject or accept $H_0$
will depend on the observed value of $B(x)$, where small values of
$B(x)$ correspond to the rejection of $H_0$. For the traditional 
frequentist the classical most powerful test of the simple hypothesis
is determined by some critical value $c$ such that

if $B(x) \le c$, reject $H_0$,

if $B(x) > c$, accept $H_0$.

\noindent
In compliance with this test, the frequentist reports 
Type I and Type II error probabilities as
$\alpha = P_0(B(X) \le c) \equiv F_0(c)$ and
$\beta = P_1(B(X) > c) \equiv 1 - F_1(c)$, 
where $F_0$ and $F_1$ are cumulative density functions of $B(X)$ under
$H_0$ and $H_1$, respectively.
For a conventional equal-tailed test with $\alpha = \beta$, the critical value
$c$ satisfies $F_0(c) \equiv 1 - F_1(c)$.

In the same way we can construct the rejection region, find the critical
values $c_1$, $c_2$ and $c_{12}$, the probabilities $\alpha$ and $\beta$ 
for the statistics 
$s_1 = \displaystyle \frac{x-N_b}{\sqrt{N_b}}$ (for ``significance'' $S_1$), 
$s_2 = \displaystyle \frac{x-N_b}{\sqrt{x}}$ (for ``significance'' $S_2$)
and
$s_{12} = \displaystyle \sqrt{x} - \sqrt{N_b}$ 
(for ``significance'' $S_{12}$).  
Here, the value of $x-N_b$ is the estimation of the number of signal events.
Note that ``significance'' $S_{12}$ depends on $S_1$ and
$S_2$, namely, 
$S_{12} = \displaystyle \frac{S_1 \cdot S_2}{S_1 + S_2}$~\cite{4}.

\section{Determination of probability density functions for statistics}

The probability density functions of statistics under consideration
can be obtained in an analytical form. Another way to obtain
the p.d.f. is the calculations by a Monte Carlo simulation 
of the results of a large number of experiments 
(see as an example \cite{7,6,8})
for the given values $N_s$ and $N_b$. 
In this study we use the latter approach.
The p.d.f.'s for $N_s+N_b=104$ and $N_b=53$ obtained by this way are shown 
in Fig.3 (these distributions are the result of $10^5$ simulation experiments 
for random variables $\xi$ and $\eta$). 
The difference between these p.d.f.'s and p.d.f.'s resulting 
from direct calculations of the probabilities (Fig.1) is extremely small. 

\begin{figure}[ht]
\epsfig{file=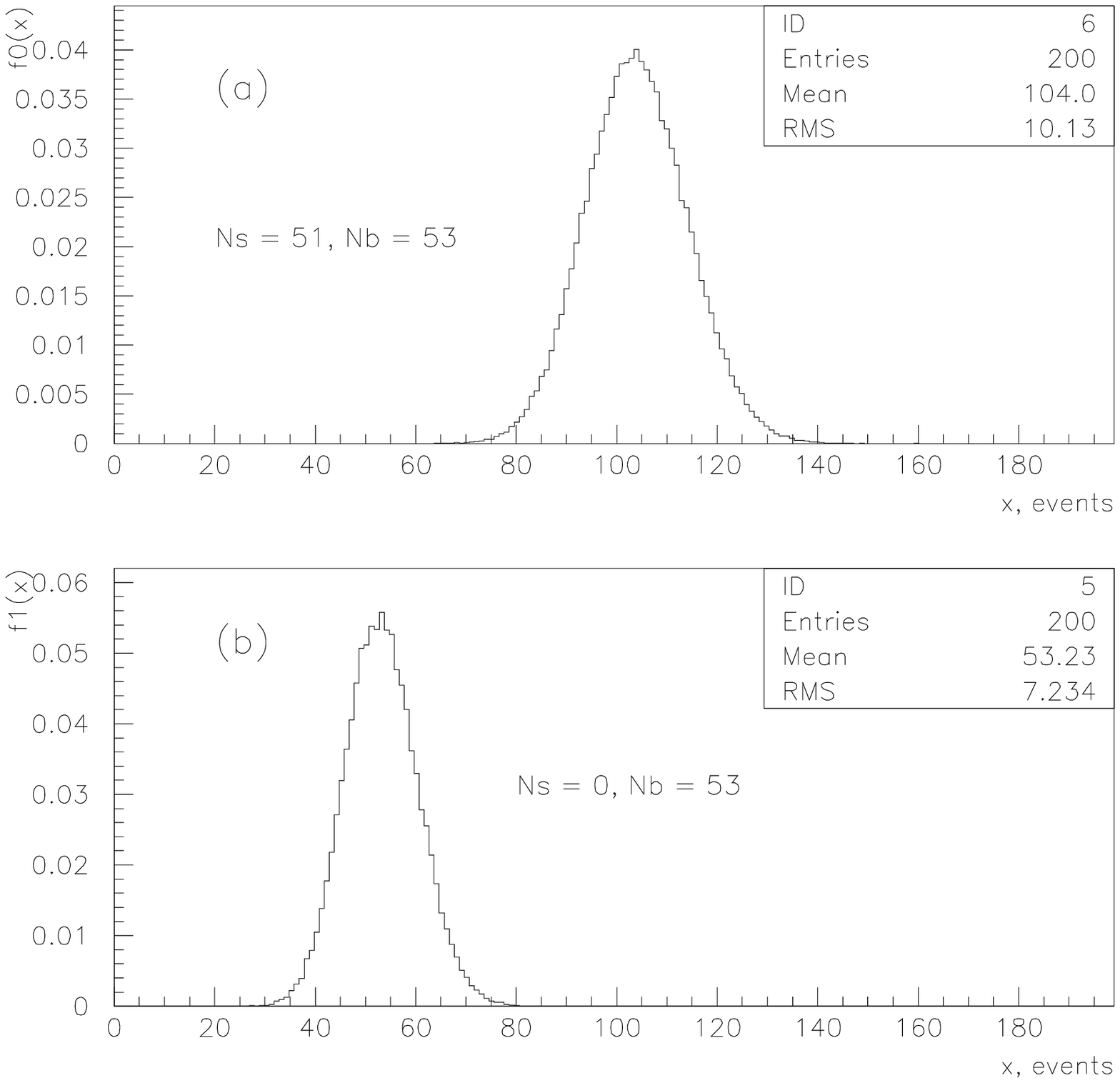,width=\textwidth}
\caption{The probability density functions
$f_0(x)$~(a) and $f_1(x)$~(b) for the case of 51 signal events 
and 53 background events obtained by Monte Carlo simulation.}
\label{fig.3}
\end{figure}

In Fig.4 the p.d.f.'s of statistic $s_2$ for the case 
of $N_s = 51$, $N_b = 53$~(a) and the case of $N_s = 0$, $N_b = 53$~(b)
are shown. The behaviour of probabilities $\alpha$ 
and $\beta$ versus the critical value $c_2$ for the statistic $s_2$ 
is also presented in Fig.4~(c). 

\begin{figure}[ht]
\epsfig{file=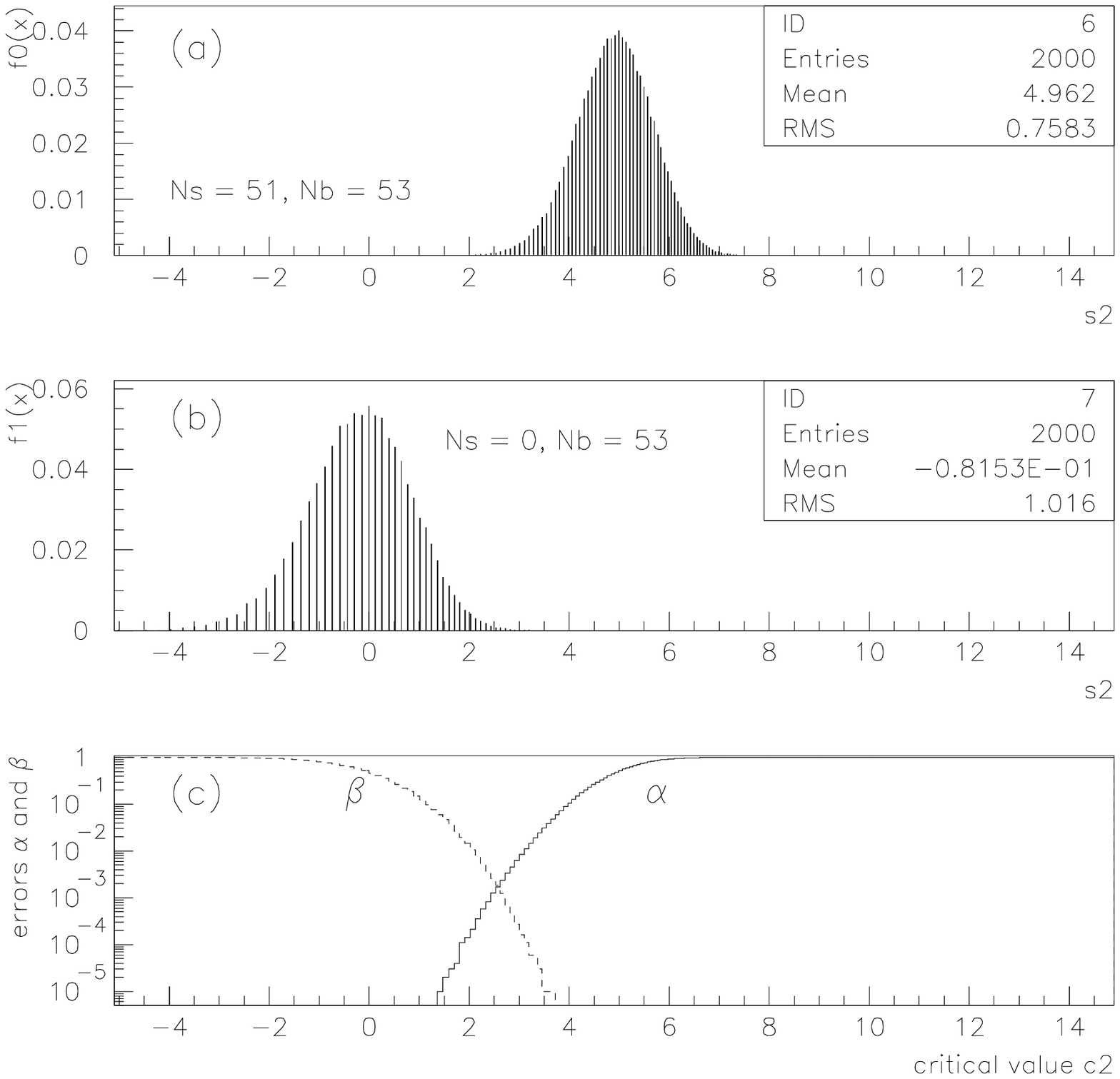,width=\textwidth}
\caption{The probability density functions $f_0(x)$~(a) and $f_1(x)$~(b) 
of statistic $s_2$. The dependence of Type I and Type II errors on
critical value $c_2$~(c) for the case of 51 signal events 
and 53 background events.}
\label{fig.4}
\end{figure}

It is worth to stress that this approach allows one to construct the p.d.f.'s
and, correspondingly, the acceptance
and the rejection regions for complicated statistics with account for the 
systematic errors and the uncertainties in $N_b$ and $N_s$ estimations.

\section{Comparison of different statistics}

We compare the statistic $s_1$, the statistic $s_2$,
the statistic $s_{12}$ and the likelihood ratio ($B(x-N_b)$ in our case). 
The reason for the comparison 
is the existence of a opinion that the value of such type statistic
$(s_1,~s_2,~s_{12})$
characterizes the difference between the samples with and without signal events 
in terms of ``standard deviations'' 
($1~\sigma,~2~\sigma, \dots,~5~\sigma$)~\footnote{If $f_1(x)$ is 
the standard normal distribution, then the $1~\sigma$ deviation
from 0 corresponds the area of tail that is equal to $0.1587$,
$2~\sigma$~--~$0.0228$, $3~\sigma$~--~$0.00135$, $4~\sigma$~--~$0.000032$
and $5~\sigma$~--~$0.000003$.}.
To anticipate a little,
the values of $\alpha$ and $\beta$ corresponding to these
``standard deviations'' depend on the value of the sample and for $S_1$,
for example, $\alpha$ and $\beta$ have a perceptible value even if
$N_s$ and $N_b$ satisfy the condition $S_1 = 5$.

The Type I error $\alpha$ is also called a significance level of the test.
The value for $\beta$ is meaningful only when it is related to an alternative
hypothesis $H_1$.
The dependence $1-\beta$ is referred to as a power function that allows one
to choose a preferable statistic for the hypothesis testing.
It means that for the specified significance level we can determine the
critical value $c$ (correspondingly, $c_1,~c_2,~c_{12})$ 
and find the power $1-\beta$ of this criterion.
The greater the value $1-\beta$, the better statistic
separates hypotheses for the specified value of $\alpha$. 

In Table 1 the comparison result is shown. For several values
of $N_s$ and $N_b$  
(significance level $\alpha = 0.01$)~\footnote{The conditions 
$min(0.01 - \alpha)$ and $\alpha \le 0.01$ are performed.}~ 
the critical values $c_1$, $c_2$, $c_{12}$, $c$ 
and the corresponding values of power $1-\beta$ of these criteria
for the statistics $s_1$, $s_2$, $s_{12}$ and the 
likelihood ratio are presented. 
As is seen from Table~I there is no visible difference in the power values
for the considered statistics,
i.e. we can use in an equivalent manner either of these statistics 
for the hypotheses testing. 

\begin{table}[h]
\small
    \caption{The comparison of power of criteria for different
statistics. The values $c_1$, $c_2$, $c_{12}$ and $c$
are the critical values of statistics $s_1$, $s_2$, $s_{12}$ and
likelihood ratio for $\alpha=0.01$. 
The values $1-\beta$ are the power for corresponding critical values.}
    \label{tab:Tab.1}
    \begin{center}
\begin{tabular}{|l|l|ll|ll|ll|ll|}
\hline
&statistic:& &$s_1$        & &$s_2$       & & $s_{12}$  &likelihood&ratio \\
\hline
$N_s$&$N_b$&$c_1$&$1-\beta$&$c_2$&$1-\beta$&$c_{12}$&$1-\beta$&$c$&$1-\beta$\\ 
\hline
  10 & 5   & 0.89 &  0.762  & 0.75&  0.762 & 0.3 & 0.762 & 0.035& 0.760\\
  15 &     & 2.23 &  0.968  & 1.58&  0.968 & 0.8 & 0.968 & 0.078& 0.968\\
  20 &     & 4.02 &  0.999  & 2.40&  0.999 & 1.4 & 0.999 & 2.563& 0.999\\
  25 &     & 5.81 &  1.000  & 3.06&  1.000 & 1.9 & 1.000 & 110.0& 1.000\\
\hline
  15 &10   & 1.26 &  0.864  & 1.06&  0.866 & 0.4 & 0.865 & 0.045& 0.864\\
  20 &     & 2.52 &  0.986  & 1.88&  0.986 & 0.9 & 0.985 & 0.269& 0.986\\
  25 &     & 3.79 &  0.999  & 2.55&  0.999 & 1.4 & 0.999 & 3.939& 0.999\\
  30 &     & 5.05 &  1.000  & 3.13&  1.000 & 1.8 & 1.000 & 307.0 & 1.000\\
\hline
  15 &15   & 0.77 &  0.750  & 0.70&  0.747 & 0.2 & 0.750 & 0.040 & 0.749\\
  20 &     & 1.80 &  0.947  & 1.49&  0.947 & 0.7 & 0.948 & 0.117 & 0.947\\
  25 &     & 2.84 &  0.994  & 2.15&  0.994 & 1.1 & 0.994 & 0.667 & 0.994\\
  30 &     & 3.87 &  0.999  & 2.73&  1.000 & 1.5 & 1.000 & 7.795 & 1.000\\
\hline
  20 &55   & 0.13 &  0.535  & 0.00&  0.479 &-0.1 & 0.483 & 0.052 & 0.536\\
  25 &     & 0.67 &  0.733  & 0.64&  0.733 & 0.2 & 0.735 & 0.049 & 0.731\\
  30 &     & 1.21 &  0.873  & 1.12&  0.874 & 0.4 & 0.843 & 0.074 & 0.873\\
  35 &     & 1.88 &  0.963  & 1.68&  0.962 & 0.7 & 0.950 & 0.231 & 0.962\\
  40 &     & 2.42 &  0.989  & 2.10&  0.988 & 1.0 & 0.988 & 0.512 & 0.989\\
  45 &     & 2.96 &  0.997  & 2.60&  0.998 & 1.3 & 0.998 & 2.894 & 0.998\\
  50 &     & 3.64 &  1.000  & 2.98&  1.000 & 1.5 & 1.000 & 9.957 & 1.000\\
\hline
\end{tabular}
    \end{center}
\end{table}

\section{Equal-tailed test}

Of concern to us is the question: What is meant by the statement that

$S_1 = \displaystyle \frac{N_s}{\sqrt{N_b}} = 5$ or 
$S_2 = \displaystyle \frac{N_s}{\sqrt{N_s + N_b}} = 5$~?\\

Tables 2 and 3 give the answer to this question. 
In Tables 2 and 3 the values $N_s$ and $N_b$ corresponding to the 
above condition, the values $\alpha$ and $\beta$ determined 
by applying equal-tailed test (in this study we use the conditions 
$min(\beta - \alpha)$ and $\alpha \le \beta$) are presented. One can see 
the dependence of $\alpha$ (or $\beta$) on the value of sample. 
The case of $N_s=5$ and $N_b=1$ for $S_1$~(Fig.5) is perhaps the most
dramatic example. We have $5 \sigma$ deviation,
however, if we reject the hypothesis $H_0$, we are mistaken in $6.2\%$ of cases
and if we accept the hypothesis $H_0$ we are mistaken in $8.0\%$ of cases.

\begin{table}[h]
\small
    \caption{The dependence of $\alpha$ and $\beta$ determined by 
using equal-tailed test on $N_s$ and $N_b$
for $S_1=5$. The $\kappa$ is the area of intersection of probability density 
functions $f_0(x)$ and $f_1(x)$.}
    \label{tab:Tab.2}
    \begin{center}
\begin{tabular}{|l|l|l|l|l|}
\hline
$N_s$ & $N_b$ & $\alpha$ & $\beta$ & $\kappa$ \\ 
\hline
   5 &  1   &   0.0620  &  0.0803 & 0.1423   \\
  10 &  4   &   0.0316  &  0.0511 & 0.0828   \\
  15 &  9   &   0.0198  &  0.0415 & 0.0564   \\
  20 & 16   &   0.0141  &  0.0367 & 0.0448   \\
  25 & 25   &   0.0162  &  0.0225 & 0.0383   \\
  30 & 36   &   0.0125  &  0.0225 & 0.0333   \\
  35 & 49   &   0.0139  &  0.0164 & 0.0303   \\
  40 & 64   &   0.0114  &  0.0171 & 0.0278   \\
  45 & 81   &   0.0124  &  0.0136 & 0.0260   \\
  50 & 100  &   0.0106  &  0.0143 & 0.0245   \\
  55 & 121  &   0.0114  &  0.0120 & 0.0234   \\
  60 & 144  &   0.0100  &  0.0126 & 0.0224   \\
  65 & 169  &   0.0106  &  0.0109 & 0.0216   \\
  70 & 196  &   0.0095  &  0.0115 & 0.0209   \\
  75 & 225  &   0.0101  &  0.0102 & 0.0203   \\
  80 & 256  &   0.0091  &  0.0107 & 0.0198   \\
  85 & 289  &   0.0096  &  0.0097 & 0.0193   \\
  90 & 324  &   0.0088  &  0.0101 & 0.0189   \\
  95 & 361  &   0.0081  &  0.0106 & 0.0185   \\
 100 & 400  &   0.0086  &  0.0097 & 0.0182   \\
 150 & 900  &   0.0078  &  0.0084 & 0.0162   \\
 500 & $10^4$ &  0.0068  &  0.0068 & 0.0136  \\
 5000 & $10^6$ &  0.0062  &  0.0065 & 0.0125 \\
\hline
\end{tabular}
    \end{center}
\end{table}

\begin{table}[h]
\small
    \caption{The dependence of $\alpha$ and $\beta$ determined by 
using equal-tailed test on $N_s$ and $N_b$ for $S_2 \approx 5$. 
The $\kappa$ is the area of intersection of probability density functions
$f_0(x)$ and $f_1(x)$.}
    \label{tab:Tab.3}
    \begin{center}
\begin{tabular}{|l|l|l|l|l|}
\hline
$N_s$ & $N_b$ &  $\alpha$ & $\beta$ & $\kappa$ \\ 
\hline
  26 &   1 &  $0.519 \cdot 10^{-5}$ & $0.102 \cdot 10^{-4}$ &  
$0.154 \cdot 10^{-4}$   \\
  29 &   4 &  $0.661 \cdot 10^{-4}$ & $0.764 \cdot 10^{-4}$ &  
$0.142 \cdot 10^{-3}$  \\
  33 &   9 &  $0.127 \cdot 10^{-3}$ & $0.439 \cdot 10^{-3}$ &  
$0.440 \cdot 10^{-3}$  \\
  37 &  16 &  $0.426 \cdot 10^{-3}$ & $0.567 \cdot 10^{-3}$ &  
$0.993 \cdot 10^{-3}$   \\
  41 &  25 &  $0.648 \cdot 10^{-3}$ & $0.118 \cdot 10^{-2}$ &  
$0.172 \cdot 10^{-2}$  \\
  45 &  36 &  $0.929 \cdot 10^{-2}$ & $0.193 \cdot 10^{-2}$ &  
$0.262 \cdot 10^{-2}$  \\
  50 &  49 &  $0.133 \cdot 10^{-2}$ & $0.185 \cdot 10^{-2}$ &  
$0.314 \cdot 10^{-2}$  \\
  55 &  64 &  $0.178 \cdot 10^{-2}$ & $0.179 \cdot 10^{-2}$ &  
$0.357 \cdot 10^{-2}$  \\
 100 & 300 &  $0.317 \cdot 10^{-2}$ & $0.428 \cdot 10^{-2}$ &  
$0.735 \cdot 10^{-2}$  \\
 150 & 750 &  $0.445 \cdot 10^{-2}$ & $0.450 \cdot 10^{-2}$ &  
$0.894 \cdot 10^{-2}$  \\
\hline
\end{tabular}
    \end{center}
\end{table}

\begin{figure}[ht]
\epsfig{file=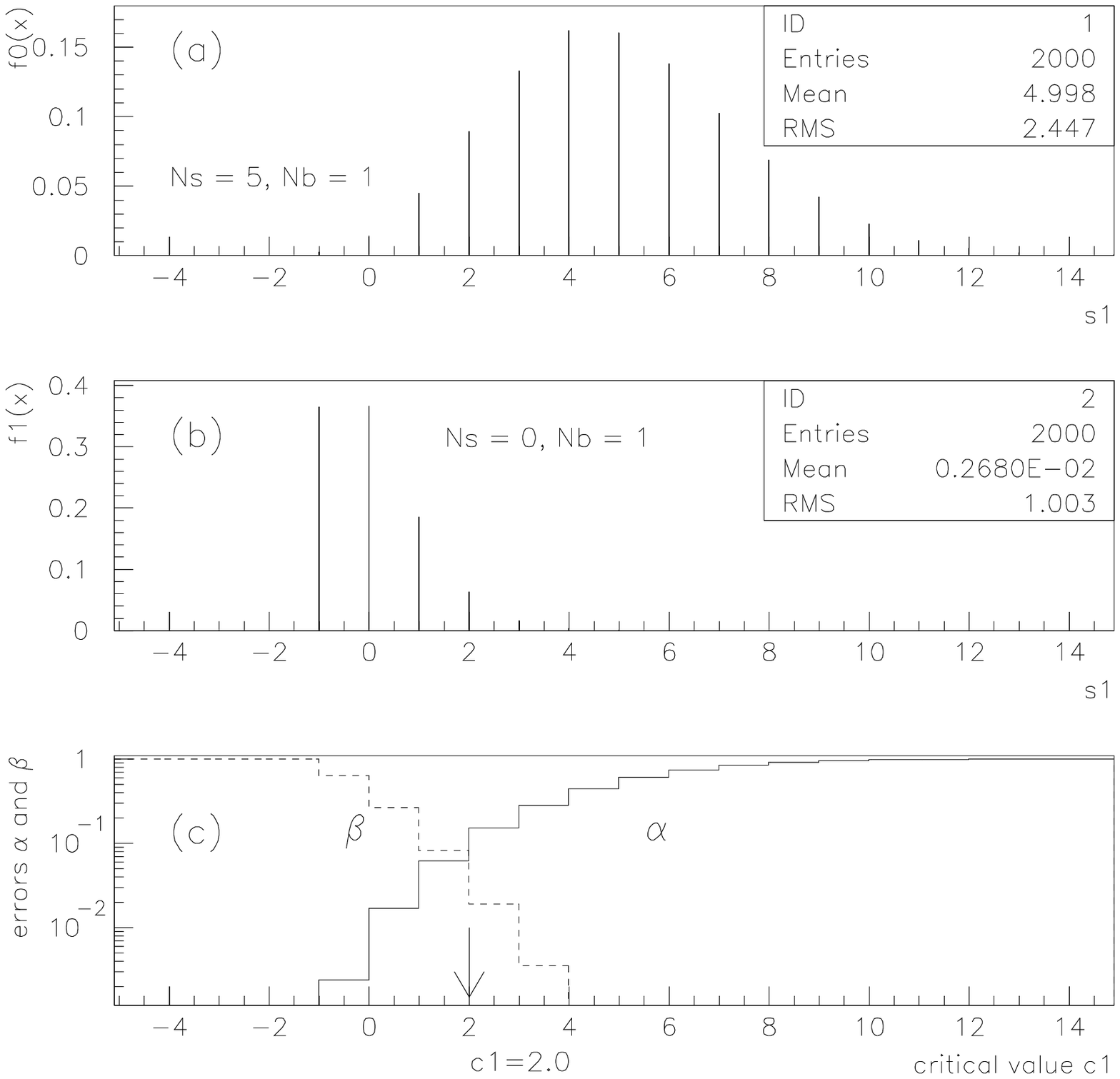,width=\textwidth}
\caption{The probability density functions $f_0(x)$~(a) and $f_1(x)$~(b) 
of statistic $s_1$. The dependence of Type I and Type II errors on
critical value $c_1$~(c) for the case of 5 signal events 
and 1 background events.}
\label{fig.5}
\end{figure}
 
One can point out that for a good deal of events the values of $\alpha$
for $S_1$ and $S_2$ approach each other. A simple argument explains
such dependence. The $x-N_b$ has the variation equal to
$\sqrt{N_s+N_b}$ for nonzero signal events, and to $\sqrt{N_b}$ if
signal events are absent. Correspondingly, if $N_b~\gg~N_s$,
the contribution of $N_s$ to the variation is very small. 
Therefore,
the standard deviation tends to unity both for the distribution
of $s_1$~(Fig.6) and for the distribution of $s_2$.
It means that for the sufficiently large $N_b$, the values of
$\alpha$ and $\beta$ obtained by equal-tailed test have a constant value close
to 0.0062. These distributions also can be approximated
by a standard Gaussian ${\cal N}(0,1)$~\footnote{It is a conventional notation
for normal distribution ${\cal N}$(mean,variance).}~ 
for the pure background and Gaussian ${\cal N}(5,1)$ 
for the signal mixed with the background. Therefore, the equal-tailed test
for the normal distributions gives $c_1 = 2.5$ and 
$\alpha = \beta = 0.0062$. These are the limiting values of $\alpha$ 
and $\beta$ for the requirement $S_1 = 5$ or $S_2 = 5$ (by the way
$S_{12}$ equals 2.5 in this case). 

\begin{figure}[ht]
\epsfig{file=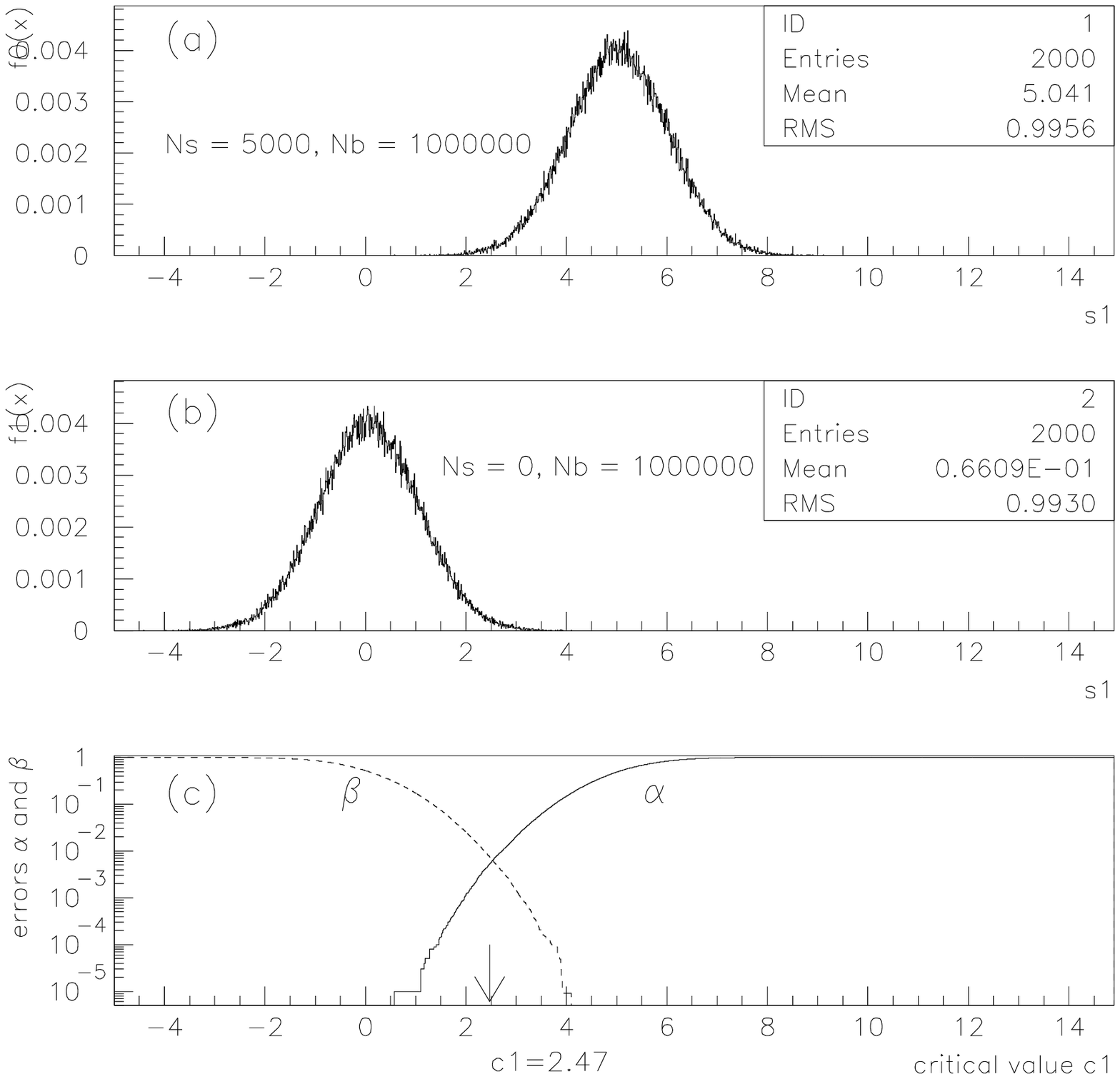,width=\textwidth}
\caption{The probability density functions $f_0(x)$~(a) and $f_1(x)$~(b) 
of statistic $s_1$. The dependence of Type I and Type II errors on
critical value $c_1$~(c) for the case of 5000 signal events 
and $10^6$ background events.}
\label{fig.6}
\end{figure}

In a similar way we can determine the behaviour of the Type I and Type II 
errors depending on $N_s$ and $N_b$
for a small number of events and we can predict the limiting values
of $\alpha$ and $\beta$
for a large number of events in case of other statements about 
statistic $s_1$~(Table 4) or any other estimator.

\begin{table}[H]
    \caption{The dependence of $\alpha$ and $\beta$ determined by 
using equal-tailed test on $N_s$ and $N_b$ for $S_1 = 2$,
$S_1 = 3$, $S_1 = 4$, $S_1 = 6$ and $S_1 = 8$. 
The $\kappa$ is the area of intersection of probability density functions
$f_0(x)$ and $f_1(x)$.}
    \label{tab:Tab.4}
    \begin{center}
\begin{tabular}{|l|l|l|l|l|l|}
\hline
$S_1$ & $N_s$ & $N_b$ & $\alpha$ & $\beta$ & $\kappa$ \\ 
\hline
2 &    2 &   1 &  0.199  &      0.265  &  0.4634  \\   
  &    4 &   4 &  0.192  &      0.216  &  0.4061  \\  
  &    6 &   9 &  0.184  &      0.199  &  0.3817  \\    
  &    8 &  16 &  0.179  &      0.188  &  0.3680  \\   
  &$\infty$&$\infty$& 0.1587 & 0.1587  &  0.3174 \\
\hline
3 &    3 &   1 &  0.0906 &     0.263 &  0.3184  \\
  &    6 &   4 &  0.0687 &     0.216 &  0.2408  \\ 
  &    9 &   9 &  0.0917 &     0.123 &  0.2159  \\
  &   12 &  16 &  0.0722 &     0.131 &  0.1952  \\
  &$\infty$&$\infty$& 0.0668 & 0.0668 & 0.1336 \\
\hline
4 &    4 &   1 &  0.0400 &     0.263 & 0.2050  \\
  &    8 &   4 &  0.0459 &     0.110 & 0.1406 \\
  &   12 &   9 &  0.0424 &     0.0735 & 0.1130  \\
  &   16 &  16 &  0.0407 &     0.0572 & 0.0977 \\
  &$\infty$&$\infty$& 0.0228 & 0.0228 & 0.0456 \\
\hline
6 &    6 &   1 &  0.0301  &    0.0806 & 0.1008   \\
  &   12 &   4 &  0.0217  &    0.0217 & 0.0434  \\
  &   18 &   9 &  0.0089  &    0.0224 & 0.0271  \\
  &   24 &  16 &  0.00751 &    0.0132 & 0.0198  \\
  &$\infty$&$\infty$& 0.00135 & 0.00135 & 0.0027 \\
\hline
8 &    8 &   1 &  0.0061   &   0.0822 & 0.0402 \\
  &   16 &   4 &  0.0049   &   0.0081 & 0.0131  \\
  &   24 &   9 &  0.0016   &   0.0052 & 0.00567 \\
  &   32 &  16 &  0.00128  &   0.00237 & 0.00331 \\
  & $\infty$ & $\infty$& 0.000032 & 0.000032 & 0.000064 \\
\hline
\end{tabular}
    \end{center}
\end{table}

Right column in Tables 2, 3 and 4 contains the value of probability
$\kappa$~\cite{4}. 
The $\kappa$ is a characteristic of the observability of Phenomenon
for the given $N_s$ and $N_b$. In particular, it is the fraction of p.d.f.
$f_0(x)$ for statistic $x$
that can be described by the fluctuation of background
in case of the absence of Phenomenon.
The value of $\kappa$ equals the area of intersection of probability density
functions $f_0(x)$ and $f_1(x)$ (Fig.1). 
Clearly, if we superimpose the p.d.f.'s $f_0(x)$ and $f_1(x)$ and choose 
the intersection point of curves 
(point $\displaystyle N_{ev} = [ \frac{N_s}{ln(1 + \frac{N_s}{N_b})}]$)
as a critical value for the 
hypotheses testing~\footnote{Notice that in this point  
$f_0(N_{ev}) = f_1(N_{ev})$ (in our case conditions
$min(f_0(N_{ev}) - f_1(N_{ev}))$ and $f_1(N_{ev}) \le f_0(N_{ev})$
are performed).
By this is meant that this checking can be named as the equal probability 
test. Of course, if we use the hypotheses testing
we can also determine $N_{ev}$ having found the minimum of the sum
of $\alpha$ and $\beta$ or having found the minimum of the 
sum of weighted $\alpha$ and $\beta$ or having exploited 
any other condition in accordance with the requirements of experiment.
The $\kappa$ may be thought of as independing of these requirements.},
we have $\kappa \equiv \alpha + \beta$. 
As is seen from Tables 2, 3 and 4 the value of $\kappa$ is also close to 
the sum $\alpha + \beta$ determined by using the equal-tailed test. 

The accuracy of determination of the critical value 
by Monte Carlo calculations depends on the number of Monte Carlo trials
and on the level of significance defined by the critical value.
To illustrate, Fig.7 shows the distribution of the estimations of 
the value $\displaystyle \frac{\alpha + \beta}{2}$
for the case $N_s=100$, $N_b=500$ and for the $10^5$ Monte Carlo trials
in each estimation (equal-tailed test is used).   
The result obtained via the direct calculations of p.d.f.'s 
is also shown in this Figure. Thus, this method is accurate enough to give
reliable results for estimation of the discovery potential of the experiment.

\begin{figure}[ht]
\epsfig{file=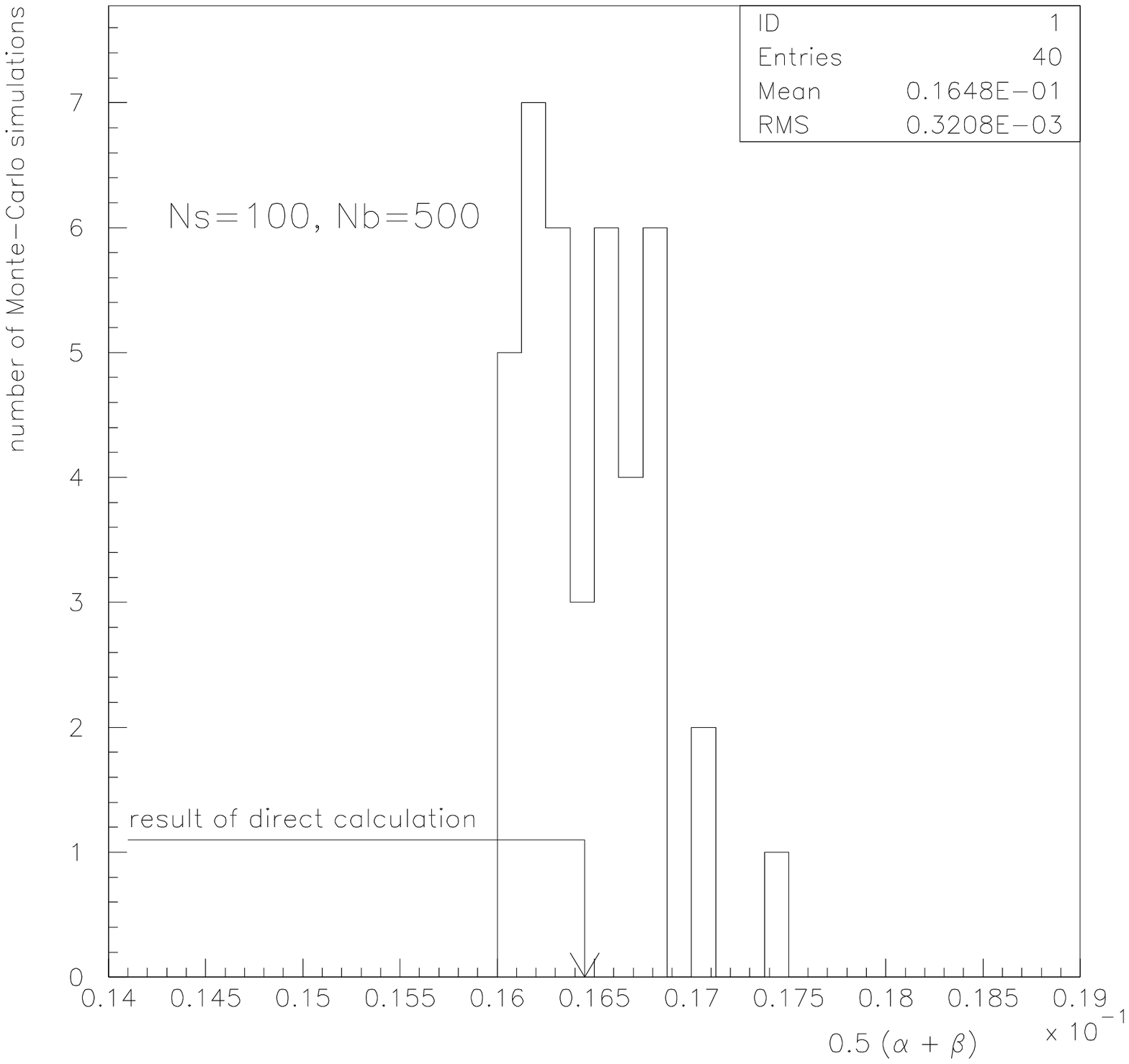,width=\textwidth}
\caption{The variation of $\displaystyle \frac{\alpha+\beta}{2}$ 
in the equal-tailed hypotheses testing 
($N_s=100$, $N_b=500$ and $N_s=0$, $N_b=500$ 
in 40 Monte Carlo simulations of probability density functions).}
\label{fig.7}
\end{figure}

The approach to the determination of the critical region in the hypotheses 
testing by Monte Carlo calculation of p.d.f.'s
can be used to estimate the integrated luminosity which is necessary 
for detection the predicted effects with sufficient accuracy. In Fig.8~(a) the 
dependence of $N_{ev}$
on integrated luminosity
(\cite{3}, Table.12, cut.5, $m_{\chi_1}=85~GeV,~N_s=45,~N_b=45$) is shown. 
The corresponding values of $\alpha$ and $\beta$ are presented in Fig.8~(b). 
As evident from Figure the integrated luminosity $L = 8 \cdot 10^4 pb^{-1}$ 
is sufficient to detect sleptons under the requirement
that the probability $\kappa \approx \alpha + \beta$ less than~$1\%$.

\begin{figure}[ht]
\epsfig{file=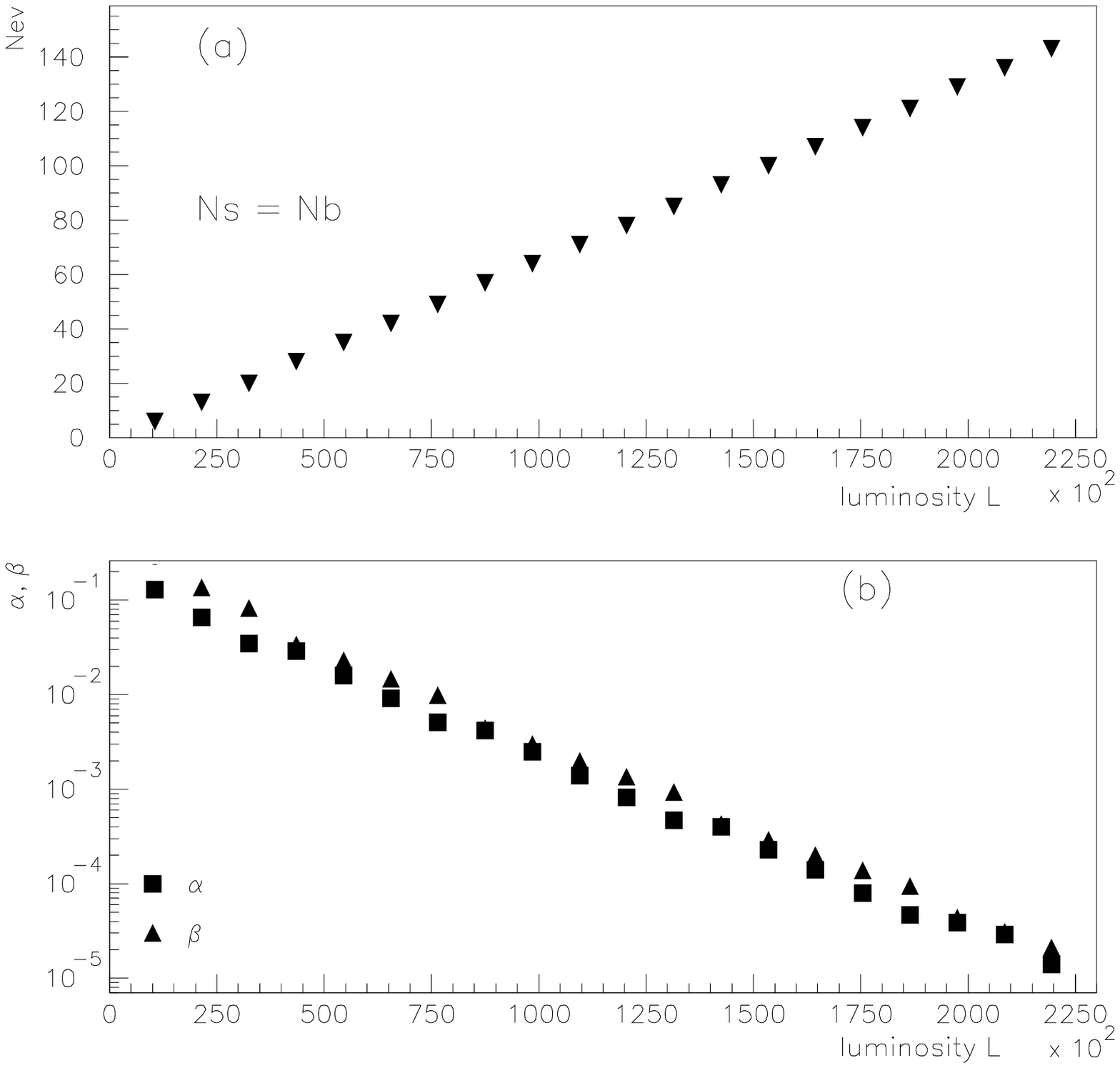,width=\textwidth}
\caption{The dependence of the critical value $N_{ev}$ (a),
Type I and Type II errors~(b) on integrated luminosity $L$ for the case
$N_s=N_b$ and $N_s=45$ for $L=10^5pb^{-1}$ (equal-tailed test).}
\label{fig.8}
\end{figure}

\begin{center}
{\large \bf Conclusion}
\end{center}

In this paper the discussion on the observation of new Phenomenon
is restricted to the testing of simple hypotheses in case of 
the predicted values 
$N_s$ and $N_b$ and the observed value $x$. As is stressed in \cite{5},
the precise hypothesis testing should not be done by forming a traditional 
confidence interval and simply checking whether or not the precise hypothesis 
is compatible with the confidence interval. A confidence interval \cite{8} is
usually of considerable importance in determining where the unknown
parameter  is likely to be, given that the alternative hypothesis is true,
but it is not useful in determining whether or not a precise null 
hypothesis is true.

To compare several statistics used for the hypotheses testing, we employ
the method that allows one to construct the rejection regions 
via the determination the probability density functions of these
statistics by Monte Carlo calculations. 
As is shown, the considered statistics have close values of power
for the specified significance level and can be used
for the hypotheses testing in an equivalent manner.
Also, it has been shown that the estimations of Type I and Type II
errors obtained by this method have a reasonable accuracy. 
The method was used to make the inferences on the
observability of some predicted phenomena.

\begin{center}
{\large \bf Acknowledgments}
\end{center}

We are indebted to M.Dittmar for useful discussions
which were one of the motivations to perform this study.
We are grateful to V.A.Matveev, V.F.Obraztsov and V.L.Solovianov 
for the interest and valuable comments. We would like to thank
E.N.Gorina for the help in preparing the article.

\end{document}